\documentclass{elsart}

\usepackage{amssymb}
\usepackage{graphicx}

\begin{document}

\begin{frontmatter}



\title{The scientific legacy of J.~P.~Elliott}


\author{P.~Van~Isacker}

\address{Grand Acc\'el\'erateur National d'Ions Lourds,
CEA/DSM--CNRS/IN2P3\\
BP~55027, F-14076 Caen Cedex 5, France}


\end{frontmatter}

James Philip Elliott, one of the towering figures of nuclear physics
of the second half of the twentieth century,
died on the 21st of October 2008.
Obituaries appeared in the British press but relatively little attention
was paid by the international scientific community to Elliott's contributions.
Given their importance, in particular for theoretical nuclear physics,
it seems appropriate to reflect here, two years after his death,
on the scientific legacy left by Phil Elliott.

Following a degree in mathematics, Phil Elliott began his scientific career in 1949,
working for a PhD in the Applied Mathematics Department of Southampton University
under the supervision of Hermann Jahn
(of the Jahn--Teller theorem in molecules~\cite{Jahn37}
which states that degenerate configurations
of non-linear molecules cannot be stable).
Jahn was an expert in the application of mathematical techniques
to atomic shell theory, developed in the 1940s by Giulio Racah~\cite{Racah40s},
and he introduced Phil Elliott to the world of
tensor operators,
coefficients of fractional parentage,
the group ${\rm U}(2\ell+1)$ and its subgroups---in short, to group theory.
Jahn had generalized group-theoretical methods
from atomic to nuclear shell theory~\cite{Jahn51},
the principal additional difficulty being that they require
the treatment of isobaric spin (or isospin as it is now called)
since one deals with a system of neutrons and protons,
as opposed to solely electrons in atoms.
In his PhD thesis Phil Elliott used the technique
of two-body coefficients of fractional parentage
to compute matrix elements of non-central forces
with applications to the $p$-shell nuclei $^6$Li, $^7$Li and $^{10}$B.
The calculations were carried out in $LS$ coupling
which was, until 1949, the accepted basis of the nuclear shell model.
The quasi-atomic model, as it was known then,
had been introduced by Elsasser
in a series of papers in the 1930s~\cite{Elsasser30s}.
Phil Elliott showed that two-body spin--orbit and tensor forces
are essential ingredients to reproduce the observed properties of $p$-shell nuclei,
as well as the one-body spin--orbit splitting
due to the interaction of the valence nucleons
with the nucleons in the $^4$He core.
An early version of this work (where the spin--orbit splitting was neglected)
appeared in 1951~\cite{Elliott51}
and the complete calculation was published in 1953~\cite{Elliott53}.
This early work had all the hallmarks of what was to come later,
in particular the complete mastery of algebraic manipulations
with a steadfast eye on the physics outcome of such calculations.

After obtaining his PhD degree,
Phil Elliott joined the Theoretical Physics Division
of the Atomic Energy Research Establishment at Harwell.
He initially worked on neutron transport problems for reactor theory
but soon switched again to nuclear structure,
attracted by Brian (later Lord) Flowers
who was at that time interested in extending
the group-theoretical methods of Racah and Jahn
from $LS$ to $jj$ coupling~\cite{Flowers52}.
Following the work of Mayer~\cite{Mayer49}
and of Haxel, Jensen and Suess~\cite{Haxel49,Suess49,Jensen49},
who provided a simple explanation of the magic numbers observed in nuclei,
it became necessary to develop
the classifications and the fractional-parentage methods
for the nuclear $j$ shells.
The Theoretical Physics Division at Harwell, headed by Brian Flowers,
was an intellectually stimulating place
with a very productive nuclear theory group led by Tony Skyrme,
which included, besides Phil Elliott,
also John Bell, Tony Lane, John Perring, Roger Phillips and John Soper.
This was the golden age of nuclear physics;
several of the finest advances in nuclear theory were made at Harwell
and Phil Elliott was at the centre of the action.
For example, together with Lane, he showed
how a constant two-body spin--orbit force
induces a variable one-body spin-orbit term in the nuclear mean field,
consistent with the data known at that time~\cite{Elliott54}.
A beautiful contribution, made together with Skyrme,
was the analysis of the separation of centre-of-mass motion from internal motion
in the context of shell-model calculations~\cite{Elliott55}.
The solution proposed by Elliott and Skyrme was elegant and simple,
and is used to this day.

Shell-model calculations with configuration mixing
(requiring the diagonalization of matrices of modest size
with a few hundreds of matrix elements)
were becoming possible
and, together with Flowers, Phil Elliott started a programme
to explore the feasibility of this approach to nuclei
at the beginning of the $sd$ shell,
just beyond $^{16}$O.
The isotopes $^{18}$F and $^{19}$F could be described
with a Gaussian central force with a single parameter
and a one-body spin--orbit term taken from $^{17}$O~\cite{Elliott55b}.
The energy spectrum and other properties of the nucleus $^{20}$Ne,
with two valence neutrons and two valence protons in the $sd$ shell,
could be calculated, following ideas of Wigner~\cite{Wigner37},
in a favoured-supermultiplet approximation,
that is, in a basis restricted to states that are totally symmetric in space
and totally antisymmetric in spin and isospin.
When mixing with other supermultiplets was included,
as a result of spin-dependent interactions and in particular the spin--orbit interaction,
the dimensions of the matrices became prohibitively large.
These and similar numerical calculations for other $sd$-shell nuclei consistently showed
that a single-$j$ shell would have been an inadequate basis
and that $s$ and $d$ orbits were inextricably mixed.
It was an important lesson
that was to be brilliantly exploited by Phil Elliott.

To appreciate the full extent of his subsequent achievement,
we need to take a step back
and consider the landscape of atomic nuclei
as they were understood in the middle of the previous century.
On the one hand, there was the recent achievement
of Mayer~\cite{Mayer49}
and of Haxel, Jensen and Suess~\cite{Haxel49,Suess49,Jensen49},
who in 1949 made a substantial breakthrough in our understanding of nuclei
by recognizing the importance of the spin--orbit coupling.
Until then, the interpretation of nuclei took place
in an $LS$-coupling basis, in analogy with what was found in atoms,
and it was assumed that the total orbital angular momentum $L$,
associated with the orbital motion of all nucleons,
and their total spin $S$ are conserved quantities during the nucleonic motion,
giving rise to several glaring discrepancies with observed nuclear properties,
not least the failure to explain the `magic numbers'
as found in all but the lightest nuclei.
A strong coupling between the orbital motion and the spin of each nucleon
destroys the $LS$ classification and leads to a $jj$-coupling basis,
where each nucleon has a well-defined angular momentum $j$
which arises from the coupling of its orbital angular momentum $\ell$ to its spin.
The hypothesis of a strong spin--orbit coupling lacked theoretical foundation
but was nevertheless immediately adopted because of its empirical success.
The ensuing model, now commonly referred to as the (spherical) nuclear shell model,
became the standard way of interpreting nuclei
and soon a wealth of mathematical techniques was developed,
allowing adaption to nuclear $j$ shells (reviewed, {\it e.g.}, in~\cite{Shalit63}).
The starting point of the nuclear shell model
is the {\em independent} motion of the individual nucleons in a $j$ shell,
giving rise to excitations which involve specific nucleons,
in other words, to so-called single-particle excitations.
A departure from this first-order image
is only possible through interactions between the nucleons
and the mixing of single-particle configurations.

Around the same time, alongside this independent-particle picture,
a very different model of the nucleus
was being developed by Bohr and Mottelson~\cite{Bohr53}.
Their starting point was to consider
the nucleus as a dense, charged liquid drop,
which was known since the 1930s
to provide a remarkably accurate description of nuclear masses~\cite{Weizsacker35,Bethe36}.
There was no reference to the individual motion of the nucleons
but excitations of a nucleus were associated with vibrations of the droplet's surface
or with its rotations in the case of a non-spherical shape.
The collective model, as it became known,
was vindicated by many observations,
especially in heavy nuclei
where rotational bands were found to be ubiquitous.

So, the middle of the twentieth century
saw the rise of two successful nuclear models
based on two different, seemingly contradictory,
views of the nucleus---single-particle {\it versus} collective.
This situation would still have been acceptable
if the models were to describe different nuclei or different nuclear states.
But it became increasingly clear
that this was not the case:
$sd$-shell nuclei such as $^{20}$Ne and $^{24}$Mg
were found to display collective characteristics
(rotational bands with energies that follow a $J(J+1)$ pattern,
with band members that are connected
by strong electric quadrupole transitions)
but could also be described with the spherical nuclear shell model.
How could this be?

The solution to this conundrum was given by Phil Elliott.
He was, first of all, familiar with the work of Racah
who had analyzed the substructure of the unitary algebra U($2\ell+1$)
which appears in the problem of many electrons
placed in a single-$\ell$ shell.
The study of the subalgebras of U($2\ell+1$)
must have seemed an esoteric topic to many at that time,
but Phil Elliott realized that some profound physical ideas
were hidden behind it.
For example, Racah had identified
(in the third of the series~\cite{Racah40s})
the orthogonal subalgebra SO($2\ell+1$) of U($2\ell+1$)
and had associated it with the `seniority number'---the
number of electrons not in pairs
coupled to orbital angular momentum $L=0$.
Secondly, Phil Elliott was aware of the existence
of an SU(3) subalgebra in Wigner's supermultiplet classification of $p$-shell nuclei
and of the suspicion that this was at the origin
of the rotational behaviour of nuclei such as $^8$Be.
It was thus natural
(but, as it usually goes, also remarkably insightful)
to study the algebraic substructure of U(6)
which is the relevant algebra for the $sd$ shell of the harmonic oscillator.
It turned out that U(6) contained,
besides a trivial ${\rm U}(1)\otimes{\rm U}(5)$,
corresponding to a separate treatment of $s$ and $d$ shells
(ruled out on the basis of numerical shell-model calculations),
the orthogonal subalgebra SO(6)
and, more importantly, also the subalgebra SU(3).
In fact, Phil Elliott proved~\cite{Elliott58a}
that the algebra of any {\em entire} harmonic-oscillator shell $p$, $sd$, $pf$, $sdg$,\dots
contains SU(3) as a subalgebra
and hence that valence nucleons, even if confined to such a shell,
may display rotational behaviour.
Furthermore, he was able to show~\cite{Elliott58b},
this time through an analysis of the algebraic substructure of SU(3) itself,
that a single one of its irreducible representations
corresponds to one (or several) intrinsic state(s)
out of which all states of one (or several) band(s) can be projected.
After the derivation of these results,
it remained to establish the physical origin of the reduction from U(6) to SU(3).
This, according to the recollections of Phil Elliott himself~\cite{Elliott99a},
became clear during a discussion with Ben Mottelson in Copenhagen in 1957.
It can indeed be shown that the isoscalar quadrupole interaction between nucleons
is responsible for the lowering of symmetry~\cite{Elliott58a,Elliott58b}.

The SU(3) model has been one of the most important breakthroughs
in nuclear physics for several reasons.
First of all, it bridged the gap between the two contrasting views of the nucleus
which were developed in parallel in the 1950s.
Furthermore, it did so by introducing the elegant and seminal concept
of spectrum generating algebras
(sometimes also called dynamical-symmetry algebras).
In the $sd$ shell, the large degeneracies of Wigner's supermultiplets
associated with ${\rm U}(6)\otimes{\rm SU}_{ST}(4)$
are reduced to lower ones
of the lower symmetry  ${\rm SU}(3)\otimes{\rm SU}_{ST}(4)$,
and this symmetry breaking occurs
because of the quadrupole interaction among the nucleons.
I also would like to stress the pioneering character of this work,
which occurred several years before similar ideas developed
in high-energy physics~\cite{Neeman61}
in the context of the eightfold way~\cite{Gellmann62}.
Phil Elliott's work is widely cited by nuclear physicists---the
two papers~\cite{Elliott58a,Elliott58b}
received close to 2000 citations---but
arguably is still not recognized at its true value by other physics communities.

The SU(3) scheme had one major drawback,
namely it presupposed Wigner's supermultiplet classification, hence $LS$ coupling
and, as such, it did not accommodate spin-dependent interactions
or a spin--orbit coupling.
Direct applications of SU(3) to nuclei
were therefore restricted to (the beginning of) the $sd$ shell.
Although Elliott's model shows
how deformed, non-spherical shapes
may arise out of the spherical shell model,
the argument is not {\it a priori} applicable
in heavy nuclei
where the spin--orbit coupling
causes a considerable rearrangement
of the single-particle levels.
Over the years several schemes have been proposed
with the aim of transposing the SU(3) scheme
to those modified situations.
One successful way of extending applications
of the SU(3) model to heavy nuclei
is based upon the concept of pseudo spin symmetry,
associated with the near-degeneracy of pseudo spin doublets
in the nuclear mean-field potential.
This degeneracy was noted by Hecht and Adler~\cite{Hecht69}
and, independently, by Arima~{\it et al.}~\cite{Arima69}
who realized that, because of the small pseudo spin--orbit splitting,
pseudo $LS$ coupling should be a reasonable starting point
in medium-mass and heavy nuclei
where $LS$ coupling becomes unacceptable.
With pseudo $LS$ coupling as a premise,
an SU(3) model can be constructed
in much the same way as Elliott's SU(3) model
can be defined in $LS$ coupling.
The ensuing pseudo SU(3) model
was investigated for the first time in~\cite{Ratna73}
with many applications following afterwards
(for a review, see~\cite{Draayer93}).
Finally, it is only many years after its original suggestion
that Ginocchio showed pseudo spin to be a symmetry
of the Dirac equation which occurs
if the scalar and vector potentials
are equal in size but opposite in sign~\cite{Ginocchio97}.
Another modification of SU(3)
has been suggested by Zuker~{\it et al.}~\cite{Zuker95}
under the name of quasi SU(3)
and starts from the observation
that in $jj$ coupling the single-particle matrix elements
of the quadrupole operator are largest for $\Delta j=0,\pm2$,
those with $\Delta j=\pm1$ being significantly smaller.
As a result, the spaces built from $j=l+{1\over2}$ and $j=l-{1\over2}$
are effectively decoupled.
Furthermore, the numerical values of the quadrupole matrix elements
do not differ too much in $LS$ and $jj$ coupling.
This argument has been invoked by Zuker {\it et al.}~\cite{Zuker95}
as an explanation of the rotational behaviour,
even in the presence of a large spin--orbit splitting.

The SU(3) model as well as the extensions discussed so far
all share the property of being confined to a single shell,
either an oscillator or a pseudo oscillator shell.
A full description of nuclear collective motion
requires correlations
that involve configurations outside a single shell.
The inclusion of excitations
into higher shells of the harmonic oscillator,
was achieved by Rosensteel and Rowe
by embedding the SU(3) algebra
into the symplectic algebra Sp(6)~\cite{Rosensteel77}.
Recent studies indicate that Sp(6) representations
might provide a physical basis
for large-scale no-core shell-model calculations~\cite{Dytrych07}.

With his work on SU(3),
Phil Elliott's reputation
was now firmly established among nuclear physicists.
After a sabbatical year in Rochester, USA,
a brief return to Harwell
and a first university appointment at Southampton University,
he was offered in 1962 a professorship at the University of Sussex in Brighton
where he remained for the rest of his scientific career.
Throughout the 1960s he continued to work on the SU(3) model.
For practical calculations it was necessary
to extend the techniques of Racah to the algebra of SU(3)
and to calculate matrix elements in this basis~\cite{Elliott63}.
Another extension was to implement
a departure from an exact $LS$ coupling in the SU(3) scheme.
This could be achieved by noting
that the one-body spin--orbit term in the SU(3) basis,
which is mostly responsible for the breaking of $LS$ coupling,
approximately leads to a new intrinsic quantum number $K=K_L+K_S$,
the sum of the spin and orbital angular momentum projections on the $z$ axis.
The inclusion of this breaking turned out to be essential
for a correct description of odd-mass $sd$-shell nuclei~\cite{Elliott68a}.
Also, a survey of the entire $sd$ shell,
interpolating between SU(3) and $jj$ coupling,
was carried out~\cite{Bouten67}.

In parallel with these extensions of the SU(3) model,
Phil Elliott embarked upon a programme
to deduce nuclear interaction matrix elements
from experimental nucleon--nucleon phase shifts.
The main objective was to transport the information contained in the phase shifts
over into the nuclear many-body problem.
Up until then, this was usually done
by assuming some parametrized form of the nucleon--nucleon interaction
and deducing its parameters from a least-squares fit to the phase shifts,
leading, for example, to the Hamada-Johnston~\cite{Hamada62}
or the Tabakin~\cite{Tabakin64} potentials.
Phil Elliott and his collaborators showed that,
provided the two-body matrix elements are expressed in an oscillator basis,
Talmi-Moshinsky brackets~\cite{Talmi52,Moshinsky59} can be used
that transform two-particle wave functions
to centre-of-mass and relative coordinates.
Since the nucleon--nucleon interaction depends only on the latter,
characterized by the relative quantum numbers $nlsj$,
the problem can be treated in each channel $^{2s+1}l_j$ separately,
and only off-diagonal matrix elements in the radial quantum number $n$ will occur.
This method was first applied
in a Born approximation to the entire nuclear potential~\cite{Elliott67}
and later, in its definite form, by introducing an auxiliary potential
in a distorted-wave Born approximation~\cite{Elliott68b}.
As a by-product, the method established an unambiguous relation
between the matrix elements in an oscillator basis
and the decomposition of a two-body interaction
into its central, symmetric and anti-symmetric spin--orbit, and tensor parts---a
connection which is used to this day in shell-model calculations
(see, {\it e.g.}, the recent example~\cite{Smirnova10}).
A number of papers followed
where the `Sussex' interaction
was applied to a variety of nuclear properties
and systems~\cite{Elliott68c,Jackson69,Dey69}.
Although sensible results were obtained
when the density of the nucleus was constrained at the known value,
it was found that the Sussex interaction
did not have the correct saturation properties
to produce this value as the equilibrium density~\cite{Dey69}.
In the subsequent years Phil Elliott and collaborators
proposed modified versions of the Sussex interaction
by incorporating effects of a hard core~\cite{Sanderson74,Dirim75,Tripathi77}
or, alternatively, by considering the addition
of a density-dependent term~\cite{Kassis81,Halkia82,Tripathi82}.
The reader should not be under the impression
that each of these papers presents
`just' another application or extension of the Sussex interaction.
On the contrary, many of them are gems.
For example, in~\cite{Jackson69} a beautiful symmetry technique
is used to reduce a very large shell-model calculation for the triton and alpha particle
to one that could be comfortably carried out at that time.

During the 1970s Phil Elliott also worked, together with Dawber,
on a major project, the writing of a monograph on
the use of symmetry in physics~\cite{Elliott79}.
The book captures the essence of Phil Elliott's approach to physics
which is to put the powerful and abstract techniques of group theory
at the service of theories and models of physics.
It has become one of the standards in the field of group theory
that is still used today by physicists of all areas.

The middle of the 1970s
saw the rise of a new collective model of the nucleus,
the interacting boson model (IBM)
of Arima and Iachello~\cite{Arima75}
(for a review, see~\cite{Iachello87a}).
It is assumed in this model that low-energy collective states in a nucleus
can be described in terms of $s$ and $d$ bosons
with angular momenta $\ell=0$ and $\ell=2$.
The bosons are interpreted as correlated (or Cooper) pairs
formed by two nucleons in the valence shell
coupled to the corresponding angular momenta.
Consequently, a nucleus is characterized by a constant total number of bosons $N$
which equals half the number of valence nucleons
(particles or holes, whichever is smaller).
Since a single boson can exist in six different states
(one $s$ and five $d$ states),
the relevant algebra of the model is U(6),
and this must have caught the attention of Phil Elliott.
In fact, the three dynamical symmetries of the IBM,
U(5), SU(3) and SO(6)~\cite{Arima76,Arima78,Arima79},
were known to Phil Elliott from his earlier shell-model work on $sd$-shell nuclei.
He also realized, however,
that while both algebras are U(6),
the physics behind them is different.
In the IBM-1, the original and simplest version of the model
where no distinction is made between neutron and proton bosons,
all states are necessarily in a symmetric representation of U(6).
This is not the case in the nuclear shell model
when fermions are considered in the $sd$ shell.
Due to the short-range attractive nature of the nuclear force,
nucleons occupy a maximally symmetric spatial configuration
but can only do so consistent with the Pauli exclusion principle.
As a result, in the $sd$ shell,
the relevant U(6) representations are not necessarily symmetric
and this renders a straightforward connection
between the IBM-1 and the nuclear shell model difficult.
A microscopic interpretation of the IBM,
that is, its justification in terms of the nuclear shell model,
is more naturally obtained by distinguishing neutron and proton bosons,
leading to the second version of the model, the IBM-2~\cite{Arima77}.
The connection can be established by associating,
for neutrons and protons separately,
the seniority quantum number $v$ in fermion space
with $2N_d$, twice the number of $d$ bosons.
This is the essential idea behind the OAI mapping~\cite{Otsuka78a,Otsuka78b}.
Its range of application can be extended
through the use of generalized seniority~\cite{Talmi71}
but remains limited to spherical or weakly deformed nuclei
(for a review, see~\cite{Iachello87b}).

Phil Elliott realized that the underlying principle of the OAI mapping
was the identification of corresponding symmetries and quantum numbers
in fermion and boson spaces.
For a better microscopic understanding of the IBM
it was thus necessary to develop versions of the model
where shell-model symmetries were naturally built in.
This led him to propose, together with White~\cite{Elliott80},
a next version of the model, known as IBM-3,
which includes a neutron--proton boson,
({\it i.e.}, a fermion pair made up of one neutron and one proton).
Since the IBM-3 includes a complete $T=1$ triplet,
with a neutron--neutron, neutron--proton and proton--proton boson,
states with good isospin can be constructed
and associated with the corresponding shell-model states.
Soon afterwards, he proposed, together with Evans~\cite{Elliott81},
the most elaborate version of the model, IBM-4,
in which the bosons were also assigned an intrinsic spin $S$.
It included the IBM-3 bosons with $S=0$ and $T=1$
and an additional set of bosons with $S=1$ and $T=0$.
Given that the orbital structure of all bosons was identical ({\it i.e.}, $s$ and $d$),
IBM-4 states could be constructed which carried the quantum numbers
of total orbital angular momentum $L$,
total spin $S$, total angular momentum $J$ and total isospin $T$,
in addition to the supermultiplet ${\rm SU}_{ST}(4)$ labels $(\lambda,\mu,\nu)$.

On a few occasions Phil Elliott applied the IBM
as a model with adjustable parameters
for describing observed nuclear properties.
He was particularly interested in so-called `mixed-symmetry' states,
that is, collective nuclear excitations
with a non-symmetric character in the neutrons and protons,
observed for the first time in deformed nuclei by Bohle {\it et al.}~\cite{Bohle84}.
Such states were predicted in IBM-2
to occur in the energy range of 2--3~MeV~\cite{Iachello84}.
Prompted by Phil Elliott's interest,
the experimental nuclear physics group
at Sussex University, led by Dennis Hamilton, began a search
and proposed mixed-symmetric candidates in vibrational nuclei,
in particular in the $N=84$ isotones~\cite{Hamilton84}
and in several $f_{7/2}$ nuclei~\cite{Eid86,Abdelaziz88}.
On the basis of systematic studies of this type,
it is now recognized that mixed-symmetry states
are the prototypes of an entire class of collective levels
(for a review, see~\cite{Heydeun}).

Another aspect of the IBM that caught the attention of Phil Elliott
was its connection with the collective model of the nucleus.
It had been shown in 1980
by several authors simultaneously~\cite{Dieperink80,Ginocchio80,Bohr80}
that a collective interpretation of the IBM
can be established by taking the `classical limit'
of large boson number, $N\rightarrow\infty$,
and the geometric equivalents of the U(5), SU(3) and SO(6) limits
had been identified as vibrational, deformed and $\gamma$ unstable.
Phil Elliott and collaborators added to this discussion
by constructing a solvable Bohr hamiltonian
that corresponds to an IBM hamiltonian
that is transitional from U(5) to SO(6)~\cite{Elliott86a}.
In addition, the notion of classical limit
was given a group-theoretical interpretation
in terms of an algebraic contraction
and, as an explicit example, the contraction of SO(6)
to the semi-direct product [R$^5$]SO(5) of the rotor with fixed deformation
was worked out~\cite{Elliott86b}.

The problem that interested Phil Elliott most
was the microscopic justification of the IBM
in terms of the nuclear shell model
and, while the first IBM-3 and IBM-4 papers~\cite{Elliott80,Elliott81}
outlined a programme in this direction,
much remained to be done to achieve this goal.
In the two articles~\cite{Halse84,Halse85}
the IBM-4 was applied to nuclei in the $sd$ shell
but most of Phil Elliott's research in the 1980s and 1990s
was devoted to the problem of the relation
of the isospin-invariant IBM-3 to the nuclear shell model.
It led to a series of papers
in which sometimes intricate group-theoretical methods
were used to arrive at this goal.
In a first study, a single-$j$ shell with neutrons and protons
was analyzed in the context of IBM-3
and a correspondence was established
with a shell-model seniority classification~\cite{Evans85}.
Subsequently, two papers appeared
clarifying the relation between IBM-2
(where $F$ spin can be a good quantum number)
and IBM-3 (where isospin is).
In the first paper~\cite{Elliott87a} it was shown
that, when neutrons and protons occupy the same shell
and both are particle-like or hole-like,
IBM-2 wave functions lack isospin symmetry;
the second paper~\cite{Elliott87b} showed, in turn,
that isospin is approximately recovered in IBM-2
when neutrons are particle-like
and protons hole-like or {\it vice versa}.
With an impeccable scientific logic,
an IBM-3 hamiltonian was then microscopically derived
from a realistic $f_{7/2}$ shell-model hamiltonian
in the first case~\cite{Thompson87},
while an IBM-2 hamiltonian was in the second case~\cite{Thompson89}.
Also, the generalization toward odd-mass nuclei
with the formulation of an isospin-invariant version
of the interacting boson--fermion model was achieved~\cite{Elliott88}---a
model that was subsequently microscopically tested
in the $f_{7/2}$ shell~\cite{Evans88}.

While these were undoubtedly nice results
which contributed to the microscopic understanding of the boson model---as
reflected in the review articles~\cite{Elliott85,Elliott90}---they
were obtained under certain assumptions that required improvement.
One assumption concerned
the derivation of the boson hamiltonian
whose one- and two-body parts
were fixed from the two- and four-nucleon shell-model systems
and kept constant for higher boson numbers.
From studies with a single kind of nucleon
it was known, however, that seniority-reduction formulas
implied a specific boson number ($N$) dependence
of the coefficients in the IBM hamiltonian.
This $N$ dependence is essentially governed
by the quasi-spin SU(2) symmetry associated with seniority.
For a system with neutrons and protons
the quasi-spin algebra is generalized to SO(5) with two labels,
seniority and `reduced isospin'
({\it i.e.}, the isospin of the nucleons not in pairs coupled to $J=0$).
This then leads to an IBM-3 hamiltonian
with coefficients that depend on the boson number $N$
{\em and} on the isospin $T$.
While the idea is simple in principle,
it is much more difficult to work out
than in the like-nucleon case
because of the complexity of the SO(5) algebra.
Nevertheless, using earlier results obtained with Hecht~\cite{Hecht85},
Phil Elliott and his collaborators
succeeded in deriving the necessary seniority reduction formulas
both in the single-$j$~\cite{Elliott92} and the multi-$j$ case~\cite{Elliott94}.
This then allowed derivation of the corresponding $(N,T)$ dependence
of the IBM-3 hamiltonian~\cite{Evans93,Evans95}
and electromagnetic operators~\cite{Elliott96a}
and the application of a microscopically justified,
fully realistic IBM-3 calculation in a multi-$j$ case~\cite{Elliott96b}.

In one of his last papers, co-authored with Evans,
Phil Elliott applied his by now famous SU(3) model
to further the microscopic understanding of the IBM~\cite{Elliott99b}.
The starting observation was another implicit assumption
in the mappings considered so far,
namely they are all based on a seniority scheme, either SU(2) or SO(5).
But it is known that seniority is badly broken in deformed nuclei
hence putting any known mapping
on a questionable basis in such nuclei.
Elliott and Evans proposed, therefore, to use SU(3)
as the connecting symmetry between fermion and boson spaces.
As in the earlier first papers on IBM-3 and IBM-4~\cite{Elliott80,Elliott81},
an outline is given in~\cite{Elliott99b}
how this deformed mapping can be achieved
by listing the relevant representations in both spaces.
It is a programme of research that surely is worth pursuing.

Phil Elliott was elected to the Royal Society in 1980
and in 1994 was awarded the Rutherford medal
and prize by the Institute of Physics.
A nuclear physics conference held at Lewes, East Sussex, in 1998
was timed to mark the 40th anniversary of the SU(3) model,
and was attended by more than 100 delegates from all over the world.
In 2002 the European Physical Society
awarded its prestigious Lise Meitner prize
jointly to Phil Elliott and Franco Iachello, of Yale University,
``for their innovative applications of group-theoretical methods
to the understanding of atomic nuclei".
He served on many advisory committees 
and was for a long time
a member of the editorial board of Nuclear Physics A.

I would like to close these pages with some personal reminiscences.
I had the privilege to hold a post-doctoral position
at Sussex University from 1985 to 1988 under the guidance of Phil Elliott.
The nuclear theory group was small,
consisting, besides Phil Elliott,
of one permanent member (Tony Evans),
a post-doctoral fellow (myself)
and two PhD students.
We held regular meetings
(sometimes together with the experimental nuclear physics group
led by Dennis Hamilton)
which were steered by Phil Elliott.
I had experienced similar meetings a few years earlier in Mexico
and I have reminisced before~\cite{Isacker01}
how they were brilliantly orchestrated by Marcos Moshinsky
who for hours on end tirelessly explained his intricate mathematical ideas
to a small group of devoted acolytes.
Phil Elliott's meetings were different
but no less formative than those of Marcos Moshinsky.
Everybody was expected at some point to contribute to the discussion
and explain his or her latest research.
The atmosphere was never aggressive
but somehow Phil Elliott succeeded in voicing his critical opinion
in a clear but gentleman-like manner,
guiding his interlocutor towards a correct solution of the problem at hand.
His standards were high.
Above anything else,
he tried to instil in his pupils respect for experimental observation:
no model, however algebraically elegant,
would get his approval if it was disconnected from experiment.

I hope to have made clear in these pages
the high esteem I hold for the scientific achievements of Phil Elliott.
Higher still is my esteem
for the unassuming modesty he kept in spite of these towering achievements
and for the impeccable scientific integrity he displayed throughout his career.

I wish to thank Achim Richter and Karlheinz Langanke
for suggesting to write this tribute.
To Bill Gelletly, Achim Richter and Igal Talmi
I am grateful for a careful reading of earlier versions of the manuscript.

\end{document}